\documentclass[review,12pt]{elsarticle}

\usepackage{float}
\usepackage{amssymb}
\usepackage{amsmath}
\usepackage[T1]{fontenc}
\usepackage[colorlinks]{hyperref}
\usepackage{amsxtra}
\usepackage{amstext}
\usepackage{amssymb}
\usepackage{graphicx}
\usepackage{amsfonts}
\usepackage{multicol}
\usepackage{setspace}

\newcommand{\R}{\mathbb R}

\newcommand{\LL}{\mathcal L}

\newcommand{\I}{\infty}

\begin{document}

\begin{frontmatter}

\title{Spatial dynamics of floral organ formation}

\author{Yuriria Cortes-Poza}
\ead{yuriria@ciencias.unam.mx}
\address{Mathematics and Mechanics Department, IIMAS-UNAM, \\ Circuito escolar, Coyoac\'{a}n 04510, Mexico City, Mexico}

\author{Pablo Padilla-Longoria}
\ead{pablo@mym.iimas.unam.mx, pp432@cam.ac.uk}
\address{Mathematics and Mechanics Department, IIMAS-UNAM, , \\ Circuito escolar, Coyoac\'{a}n 04510, Mexico City, Mexico}

\author{Elena Alvarez-Buylla}
\ead{eabuylla@gmail.com}
\address{Functional Ecology Department, Ecology Institute-UNAM}



\maketitle

\begin{abstract}
Understanding the emergence of biological structures and their changes is a complex problem. On a biochemical level, it is based on gene regulatory networks (GRN) consisting on interactions between the genes responsible for cell differentiation and coupled in a greater scale with external factors.
In this work we provide a systematic methodological framework to construct Waddington's epigenetic landscape of the GRN involved in cellular determination during the early stages of development of angiosperms. As a specific example we consider the flower of the plant \textit{Arabidopsis thaliana}. Our model, which is based on experimental data, recovers accurately the spatial configuration of the flower during cell fate determination, not only for the wild type, but for its homeotic mutants as well. The method developed in this project is general enough to be used in the study of the relationship between genotype-phenotype in other living organisms.
\end{abstract}


\end{frontmatter}


\section{Introduction}
\label{sec:intro}
\begin{figure}[H]
\centering
  \includegraphics[width=8cm]{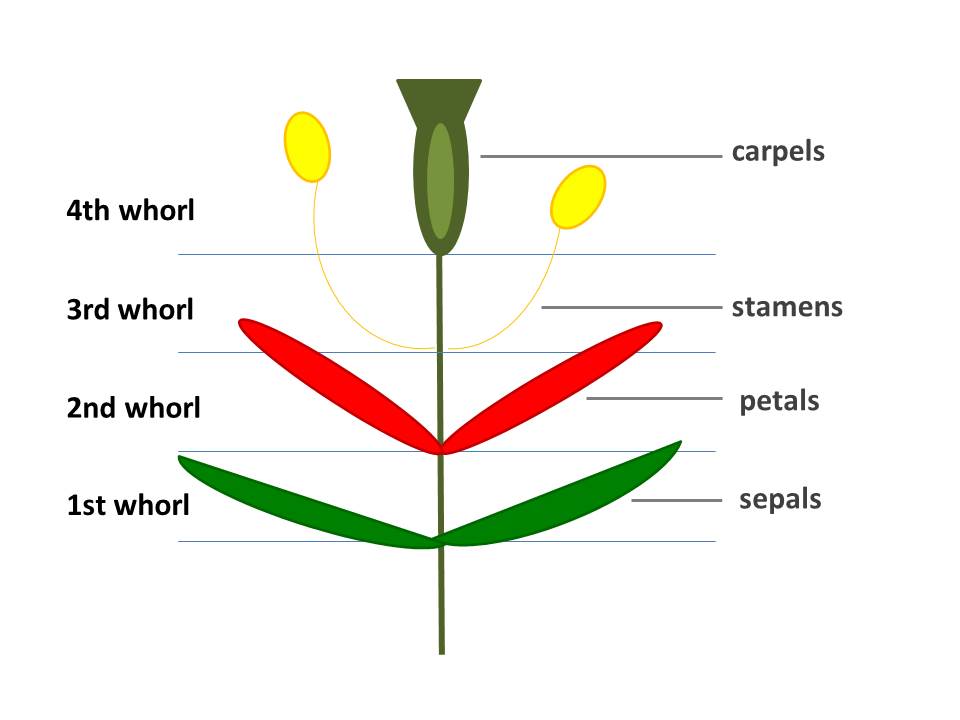}\\
  \caption{Whorls of a typical Angiosperm}\label{fig:whorls}
\end{figure}

The flower organs of all species of Angiosperms (approx. 250,000) are organized in four concentric rings (named whorls), which are, from the outer rim to the center: sepals, petals, stamens and carpels (fig (\ref{fig:whorls})). The only known exception to this configuration is the one observed in the flower \textit{Lacandonia schismatica} where the position of its stamens and carpels is inverted.

We work with the flower of the plant \textit{Arabidopsis thaliana}. This plant was the first one whose complete genome was sequenced and has been extensively studied (\cite{Alvarez-Arabidopsis}). In this paper we build and solve a mathematical model that correctly reproduces the spatial configuration  of the flower's organs in nature, during cell fate determination, that is, sepals are located in the outermost rim of the flower, then petals, stamens and carpels in the center.

In (\cite{Alvarez-FM}), we obtained the gene regulatory network (GRN) of the flower \textit{Arabidopsis thaliana}, that determines the fate of floral organ cells, using experimental data. Using this model, we construct a system of reaction-diffusion equations governed by a potential field which is the epigenetic landscape of the flower's organ formation. The epigenetic landscape models how the different environmental and genetic forces affect cellular differentiation.

We solve our system of reaction diffusion equations,  and observe that the model reproduces correctly the spatial configuration of the formation of the flower's organs. To validate our model we repeat the procedure with the homeotic mutants of the flower, which have each a different gene regulatory network (some have missing organs) and thus, a different spatial configuration. The obtained results are in agreement with experimental data.

This paper is organized in the following manner: In the first section we present some background information, including the definition of homeotic mutants, and the characteristics of those of the flower \textit{Arabidopsis thaliana} and we detail the discrete dynamical system we are basing our work on. In section two we construct our model taking into account experimental data (from the discrete dynamical system) and explain how each part is structured: starting from the definition of our model and the construction of the epigenetic landscape . In section three we explain how homeotic mutants are used to validate our model, in four we present some details of how the solution of our system is found and finally in the last section we present concluding remarks.

\section{Background information}
\label{sec:1}

\subsection{Homeotic mutants}
\label{sec:1.1}
\begin{figure}[H]
\centering
  \includegraphics[width=8cm]{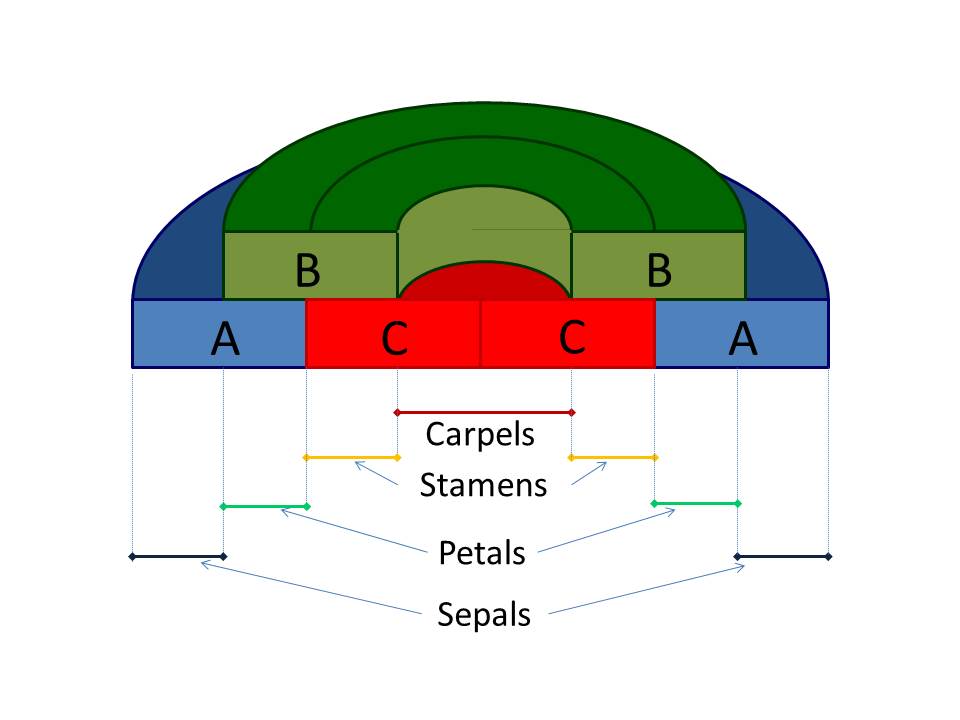}\\
  \caption{ABC model: active genes in each of the four whorls}\label{fig:abc}
\end{figure}

As mentioned in the introduction, we use homeotic mutants to validate our model. Homeotic genes are responsible for the development of specific structures in plants and animals. Mutations of these genes may cause an organ to be replaced by another. Studying these mutations, Coen and Meyerowitz in 1991 formulated the ABC model for flower development (\cite{Coen-ABC}). Even though they developed their model for the flower \textit{Arabidopsis thaliana} it is applicable to other angiosperms. According to this model, the identity of the flower organs is determined by three classes of genes: A, B and C. These classes code the transcriptional factors that in combination cause the specialization of the tissues in their specific regions during development. When flower development begins, the meristematic cells are already divided in four concentric rings (whorls). In the two outermost whorls, genes A and B are active, B genes are active in the second and third whorl and C genes are active in the third and fourth whorl (center) (see Fig.\ref{fig:abc}) (\cite{Sullivan-ABC}).
The activity and the interaction of the three types of homeotic genes determine the identity of the four flower organs. Each one of these includes the following genes:
\begin{itemize}
  \item[A)] Apetala 1 and Apetala 2
  \item[B)] Apetala 3 and Pistillata
  \item[C)] Agamous
\end{itemize}

\begin{figure}[H]
\centering
  \includegraphics[width=12cm]{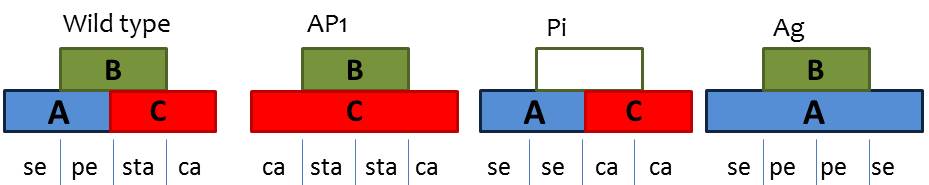}\\
  \caption{ABC model for the wild type flower and 3 mutants: AP1, Pi, Ag (shown in that order)}\label{fig:mutants}
\end{figure}

Genes A, B and C are necessary for the organ determination in the following combination: sepals: A, petals: A and B, stamens: B and C and Carpels: C. A series of mutants of the flower \textit{Arabidopsis thaliana} have been characterized. These serve as experimental material in laboratory studies. In particular we work with the mutants Apetala (AP1), Pistillata (Pi) and Agamous(Ag). The mutant AP1 has carpels in the first whorl, stamens in the second and third one and carpels in the fourth one. This mutant lacks A activity which gives rise to an expansion of the C genes in all the flower. The mutant Pi has sepals in the first and second whorl and carpels in the third and fourth one. This mutant does not have type B genes. The mutant Ag has sepals in the first whorl and then petals in next two whorls. This pattern repeats itself in the inner whorl. This mutant lacks C activity which causes an expansion of genes A. (See Fig.(\ref{fig:mutants})). These mutants have each a different gene regulatory network and each is tested on our model.

\subsection{Boolean network}
\label{sec:1.2}
In (\cite{Alvarez-GRN}) a discrete dynamical system was used to explore the dynamics of cell fate determination during the early stages of flower development. The system is a Boolean gene regulatory network, consisting of $13$ nodes (each one corresponding to a specific gene) whose state (0 or 1) is updated according to experimentally obtained rules, that correspond to the interaction between genes.

There are $2^{13}$ initial conditions. The system is iterated, starting from each initial condition. It converges to ten different attractors, each one representing one of the main cell types observed during the early stages of flower development (the meristematic cells of the inflorescence and the primordial cells of the flower meristems of sepals, petals, stamens and carpels). Each equilibrium point has thirteen components(\ref{tab:1}).

\begin{table}
\caption{Equilibrium points}
\centering
\label{tab:1}
\begin{tabular}{|l|c|}
\hline
 \multicolumn{2}{|l|}{Floral organ\qquad\qquad\qquad\ \ \ Atractor} \\
 \hline
   Inflorescence 1         & $q_1=[0,0,0,1,0,0,0,0,1,0,0,0,0]$\\
   Inflorescence 2         & $q_2=[0,0,0,1,0,0,0,0,1,0,0,0,1]$\\
   Inflorescence 3         & $q_3=[0,0,0,1,0,0,1,0,1,0,0,0,0]$\\
   Inflorescence 4         & $q_4=[0,0,0,1,0,0,1,0,1,0,0,0,1]$\\
   Sepals                  & $q_5=[0,1,1,0,1,1,0,0,0,0,1,0,0]$\\
   Petals (without UFO)    & $q_6=[0,1,1,0,1,1,0,0,0,1,1,1,0]$\\
   Petals (with UFO)       & $q_7=[0,1,1,0,1,1,0,0,0,1,1,1,1]$\\
   Stamens (without UFO)   & $q_8=[1,1,0,0,1,1,0,1,0,1,1,1,0]$\\
   Stamens (with UFO)      & $q_9=[1,1,0,0,1,1,0,1,0,1,1,1,1]$\\
   Carpels                 & $q_{10}=[1,1,0,0,1,1,0,1,0,1,1,0,0]$\\
   \hline
    \end{tabular}
\end{table}
\ \newline

Of the fixed points, we are interested only in those that correspond to the flower organs: sepals, carpels, petals and stamens, so the rest (related to vegetative organs) will not be taken into account.

It will also be of importance to count how many of the initial conditions land in each fixed point, obtaining the following information:
\begin{equation}\label{eq:bn1}
   c_S=152,c_P=160,c_T=3744,c_C=3608,
\end{equation}

where the subscripts S, P, E and C correspond to sepals, petals, stamens and carpels respectively (\cite{Espinosa-GRN},\cite{Alvarez-FM}).

Based on this discrete model, we construct a reaction-diffusion system as we explain in the following section.

\section{The model}
\label{sec:2}
We now explain how our model was constructed following two major steps: We start by building the epigenetic landscape of the flower (a potential field) that uses significant biological information, obtained from the discrete dynamical system. Then, following Turing's proposal (\cite{Turing-Morfo}), we define our model as a system of reaction-diffusion partial differential equations, which will be governed by the potential field.

\subsection{Epigenetic landscape}
\label{sec:2.2}

Epigenetic landscapes, originally proposed by Waddington in 1975 (\cite{Waddington-EL}), are developmental models that illustrate the mechanics of cell fate differentiation. These models use as a metaphor a mass in a potential field with a certain number of basins of attraction and paths that lead to each of them.
We will model the epigenetic landscape of the flower, as a potential field with four different basins of attraction, each one corresponding to a different flower organ (sepals, petals, stamens and carpels). We use the information obtained from de discrete dynamical system (\ref{sec:1.2}) to place the center and determine the size of each basin in such a way it is in correspondence with the experimental data.
For this, we recall that each stable state in the discrete dynamical system is a string with 13 characters, where each one corresponds to a specific gene of the organ, that can be off or on ($0$ or $1$ respectively). We want to place each stable state as a point on a two-dimensional plane, making similar strings (i.e. those who share several components with the same value) closer and strings with a larger number of different components further apart. That is, we want the distance between the points in the two-dimensional plane to reflect the similarity of the stable states.
For the size of each basin, we will use the reciprocal of the number of initial conditions that land in each steady state of the dynamical system, guaranteeing that equilibrium points that are reached more often will have larger basins and conversely equilibrium points that are reached fewer times will have a smaller basin.

We define the potential field on the plane $(u,v)$ determined by the epigenetic landscape in the following way:
\begin{equation}\label{eq:el1}
\begin{array}{ll}
F(u,v)=\min\{& a_S[(u-u_S)^2+(v-v_S)^2],a_P[(u-u_P)^2+(v-v_P)^2],\\
             & a_T[(u-u_T)^2+(v-v_T)^2],a_C[(u-u_C)^2+(v-v_C)^2] \},
\end{array}
\end{equation}

where $(u_S,v_S)$, $(u_P,v_P)$, $(u_T,v_T)$ and $(u_C, v_C)$ will be the centers of the basins, each one corresponding to a different flower organ (sepals, petals, stamens, carpels) and $a_S=1/c_S, a_P=1/c_P, a_T=1/c_T, c_C=1/c_C$  (where $c_S,c_T,c_P$ and $c_C$ are given in (\ref{eq:bn1}), define the size of each basin.

To compute the centers of the basins we find a plane in $\R^2$ that minimizes the square of the sum of the Euclidian distances between each point (stable state of the discrete dynamical system) and the plane. We then project each point onto, and obtain the following four points in $\mathbb{R}^2$, each corresponding to a floral organ (\ref{tab:2}).	
\begin{table}
\caption{Equilibrium points in the plane}
\centering
\label{tab:2}
\begin{tabular}{|c|l|}
\hline
 \multicolumn{2}{|l|}{Organ\qquad Fixed point} \\
 \hline
Sepals    & $p_S=(u_S,v_S)=(-1.8048, 1.0278)$ \\
Petals    & $p_P=(u_P,v_P)=(-2.5911, 0.8850)$ \\
Stamens   & $p_T=(u_T,v_T)=(-2.8466, -0.7537)$\\
Carpels   & $p_C=(u_C,v_C)=(-2.3893,-0.8381 )$\\
\hline
\end{tabular}
\end{table}
\ \newline

\begin{figure}[H]
\centering
  \includegraphics[width=10cm]{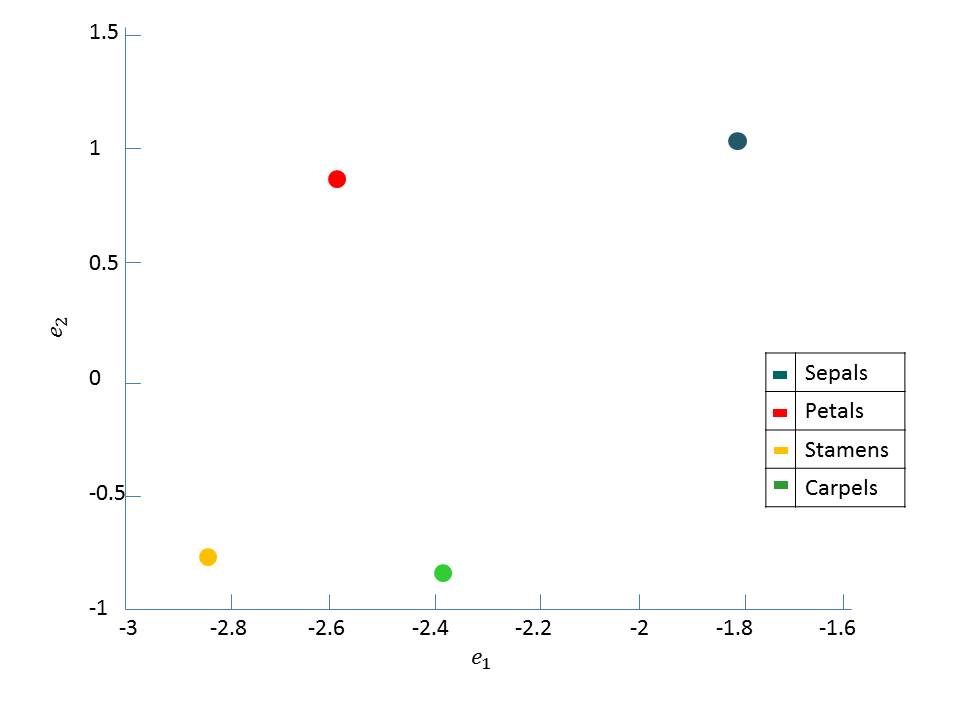}\\
  \caption{Fixed points in the plane, each corresponding to a specific floral organ}\label{conUFO}
\end{figure}

The graph of the four vectors in the plane is observed in figure (\ref{conUFO}) and the details of the computations are given in (\ref{ap:cr}).

Note that if we wish to stay in the positive octant, it suffices to do a translation in such a way that the four initial conditions
($(u_S,v_S),(u_P,v_P),(u_T,v_T)$ and $(u_C,v_C)$) are all positive.

With this information we have completed our model of the epigenetic landscape.

\subsection{Reaction diffusion system}
\label{sec:2.3}
Reaction diffusion systems are mathematical models that describe how one or more substances, distributed in the space, change, under the influence of two processes: local chemical reactions in which this substances transform each other and diffusion that makes the substances disperse. The result of this process is a stable configuration in which the chemical composition is not uniform in the spatial domain. Since 1952, when Alan Turing proposed these systems (\cite{Turing-Morfo}), they have been used to model several biological processes where pattern formation takes place.

Following Turing's ideas, our model, will be a system of reaction-diffusion partial differential equations, governed by the potential field (epigenetic landscape) $F(u,v)$ defined in the previous section (\ref{eq:el1}).

\begin{equation}\label{eq:rd1}
\begin{array}{lll}
  \frac{\partial u}{\partial t}& = & d_1\Delta u+f(u,v)\\
  \frac{\partial v}{\partial t}& = & d_2\Delta v+g(u,v)
\end{array}
\end{equation}
where $(f,g)=-\nabla F$, and $d_1,d_2$ are the diffusion constants.

The variables $u$ and $v$ represent a linear combination of the activation states of the genes, which result from choosing a system of coordinate axis in the adjusted two-dimensional plane.

The objective is to find the stationary solutions of (\ref{eq:rd1}), that is

\begin{equation}\label{rd2}
\begin{array}{lll}
  d_1\Delta u+f(u,v) & = & 0\\
  d_2\Delta v+g(u,v) & = & 0.
\end{array}
\end{equation}

Using Sturm-Liouville theory we reduce the problem to Bessel's equation, which we solve using Frobenius method. The computation of the solutions can be found in (\ref{ap:sol}).

\section{Validation of our model using the homeotic mutants}
\label{sec:3}
For each homeotic mutant (AP1,Pi and Ag), we repeat exactly the same procedure as the one used for the wild type flower: once we have the outcome of the discrete dynamical system (the equilibrium points and the number of initial conditions that land in each one of these), we compute the center and size of each basins in the potential field (epigenetic landscape). We obtain the information shown in table (\ref{tab:3}).

\begin{table}
\caption{Basins of attraction}
\centering
\label{tab:3}
 \begin{tabular}{|l|l|l|l|}
  \hline
  Mutant & Organs & Basin size & Basin center\\
  \hline
  AP1 & stamens & $c_T=1792$ & $(u_T,v_T)=(-2.9328,0.6315)$ \\
       & carpels & $c_C=1744$ & $(u_C,v_C)=(-2.5436, -0.7281)$\\
  \hline
  Pi   & sepals  & $c_S=80$ & $(u_S,v_S)=(-1.9580,-1.0798)$ \\
       & carpels & $c_C=1872$ & $(u_C,v_C)=(-2.5078, 0.8431)$ \\
  \hline
  Ag   & sepals  & $c_S=968$ & $(u_S,v_S)=(-2.0436, -0.9076)$\\
       & petals  & $c_P=992$ & $(u_P,v_P)=(-2.7466, 0.6753)$ \\
  \hline
\end{tabular}
\end{table}

 Since (as we can appreciate) the three mutants that we are studying have only two different organs each, their corresponding potential field (epigenetic landscape) will only have two basins. The system of reaction diffusion-equations of each mutant will remain unchanged (\ref{eq:rd1}).

We now present the results obtained for the wild type flower and its homeotic mutants.

\section{Results}
\label{sec:4}
\begin{figure}[H]
\centering
  \includegraphics[width=.9\textwidth]{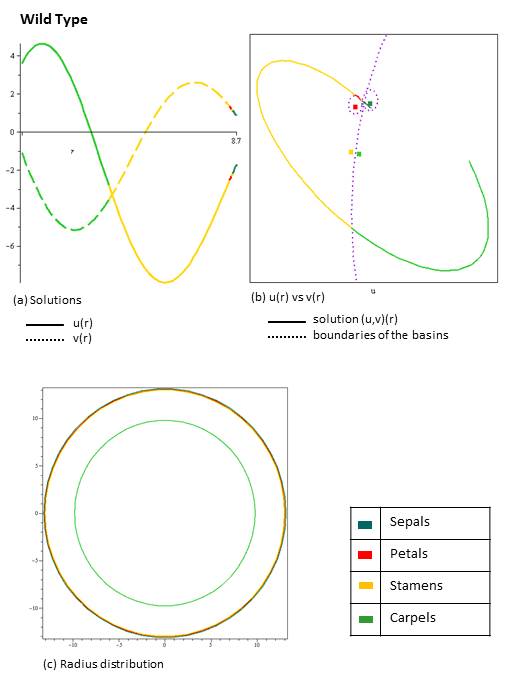}
  \caption{Results for Wild Type flower. (a) Solid line is $u(r)$ and dashed line is $v(r)$. The color of the line indicates in what basin (sepals, petals, stamens or carpels) the solution is at. (b) Solid line represents the solution $(u,v)(r)$. The color of the line indicates the basin the solution is at. The dotted line represents the boundary of the basin. (c) Radius distribution graph, it indicates the portion of the domain that each solution is comprising.}\label{fig:resWT}
\end{figure}

The solutions to the system of reaction-diffusion equations pass through the four basins of attraction in the correct order. In figure (\ref{fig:resWT}a) we show both solutions $u(r)$ and $v(r)$. The colors of the graphs mark in which basin the solution is in each value of $r$. We can see that both solutions start in the sepal's basin (at $r=R$), then go to petals, stamens and end up in carpels. We also graph the phase plane by plotting the values of the solutions $u(r)$ versus $v(r)$ in the plane u-v (figure\ref{fig:resWT}b), the dotted lines show the contour of each basin and the color dots, their centers. Finally, since the first two basins (stamens and carpels) are very small, they occupy a very small portion of the flower. We observe this phenomenon on figure(\ref{fig:resWT}c), where each color represents the portion of radius that each part of the solution (organ) takes.

Our results agree with the spatial distribution observed in the great majority of angiosperms. They suggest that the spatial pattern emerges as a result of interactions between the genes in the gene regulatory network and the action of the diffusive field.

\begin{figure}[H]
\centering
  \includegraphics[width=12cm]{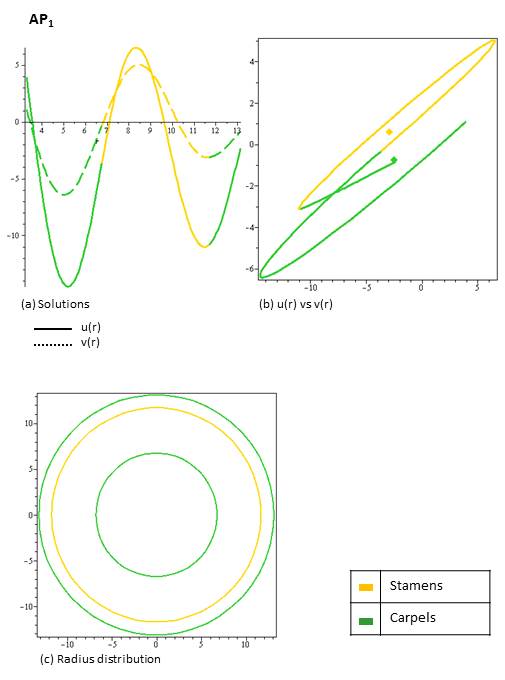}\\
  \caption{Results for mutant AP1. a) Solid line is $u(r)$ and dashed line is $v(r)$. The color of the line indicates in what basin (stamens or carpels) the solution is at. b) Solid line represents the solution $(u,v)(r)$. The color of the line indicates the basin the solution is at. The dotted line represents the boundary of the basin. c) Radius distribution graph, it indicates the portion of the domain that each solution is comprising.}\label{fig:resApt1}
\end{figure}

\begin{figure}[H]
\centering
  \includegraphics[width=12cm]{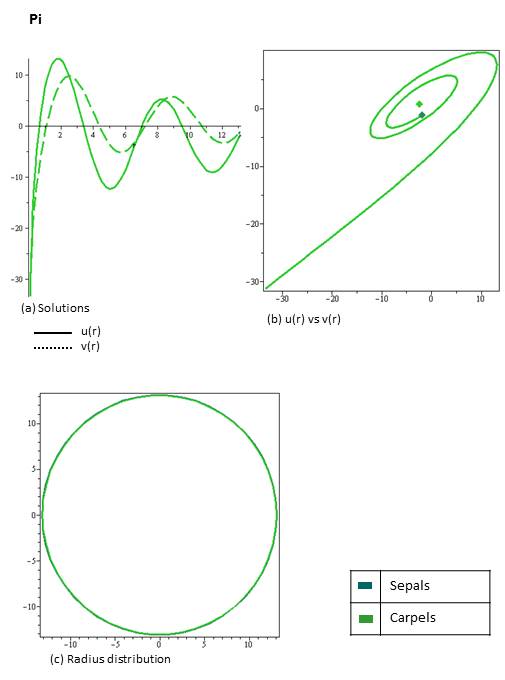}\\
  \caption{Results for mutant Pi. a) Solid line is $u(r)$ and dashed line is $v(r)$. The color of the line indicates in what basin (sepals or carpels) the solution is at. b) Solid line represents the solution $(u,v)(r)$. The color of the line indicates the basin the solution is at. The dotted line represents the boundary of the basin. c) Radius distribution graph, it indicates the portion of the domain that each solution is comprising.}\label{fig:resPi}
\end{figure}

\begin{figure}[H]
\centering
  \includegraphics[width=12cm]{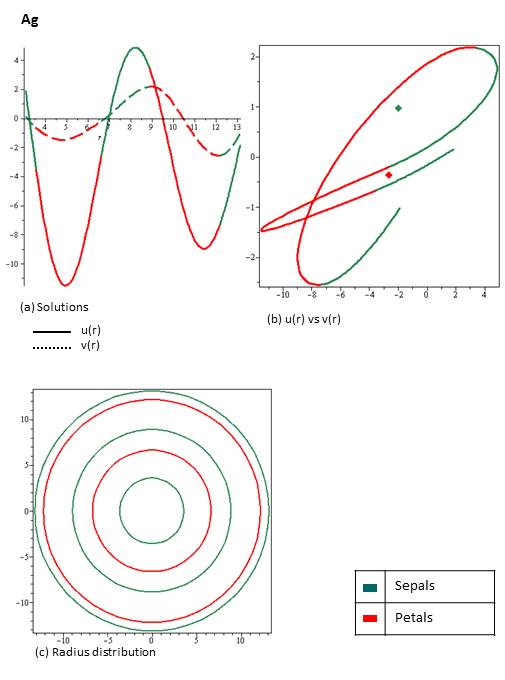}\\
  \caption{Results for mutant Ag.  a) Solid line is $u(r)$ and dashed line is $v(r)$. The color of the line indicates in what basin (sepals or petals) the solution is at. b) Solid line represents the solution $(u,v)(r)$. The color of the line indicates the basin the solution is at. The dotted line represents the boundary of the basin. c) Radius distribution graph, it indicates the portion of the domain that each solution is comprising.}\label{fig:resAg}
\end{figure}

The homeotic mutants have each a different gene regulatory network. By repeating the procedure with them, we again obtain the correct spatial distribution. In the following graphics we show these results for the mutants AP1 (\ref{fig:resApt1}), Pi (\ref{fig:resPi}) and Ag (\ref{fig:resAg}), plotting (as in the case of the wild type flower) the graph of $u$ and $v$ vs $r$ (a), then the phase plane $(u(r),v(r))$ (b) and finally the portion of radius that each organ takes up (c).

We observe that in each of these three cases the model works correctly, recovering their spatial configuration.

\section{Concluding remarks}
\label{sec:concl}
By constructing a continuous dynamical system (reaction-diffusion equations) we were able to model correctly the process of cell fate determination in the flower organs of the plant \textit{Arabidopsis thaliana} during cellular differentiation. The system of reaction-diffusion equations is solved using analytical and numerical techniques and doing it, we observe that the solutions found, recover the spatial configuration of the flower organs that appear in the analyzed flower. The model also reproduces the experimental results observed in homeotic mutants, and has the advantage of being based on a discrete system that was built using detailed experimental data.

Now that the model has been built and validated, the next goal is to take into account other physical and biological effects such as active transport, mechanical considerations and a non-trivial geometry. We would also like to take into account a growing domain and analyze the formation of biological structures in this setting.

\appendix

\section{Epigenetic landscape}\label{ap:cr}

Let $e_1$ and $e_2$ be two orthonormal vectors in $\mathbb R^{13}$ and $\Pi=<e_1,e_2>$ the plane generated by these vectors. We want to find $e_1$ and $e_2$ such that the sum of the distances of each vector (fixed point) $q_1, q_2, q_3, q_4$ to the plane $\Pi$ is the least one. We have to minimize the quantity
\begin{equation}\label{rc1}
    S=\sum_i d^2(q_i,\Pi)
\end{equation}
where $S:=S(e_1,e_2)$, and $d^2(q_i,\Pi)$ is the square of the distance of the vector $q_i$ to the plane $\Pi$. That is,
\begin{equation}\label{rc2}
    d^2(q_i,\Pi)=||q_i-Pq_i||^2,
\end{equation}
where $Pq_i$ is the projection of the vector $q_i$ to the plane $\Pi$,
\begin{equation}\label{rc3}
    Pq_i=(q_i\cdot e_1)e_1+(q_i\cdot e_2)e_2,
\end{equation}
and $(\cdot)$ denotes the scalar product between the corresponding vectors.
To minimize $S$ we use singular value decomposition (SVD), which is an excellent tool when working with sets of equations of matrices that are singular or numerically nearly singular.
We start by adjusting a plane to the set of vectors $\{q_1,q_2,q_3,q_4\}$.
Let $Q$ be the matrix of size $m\times n$ $(m=13, n=4)$, formed by placing the vectors $q_i$ as columns.

Using SVD we can decompose the matrix $Q$ in three factors and find its singular values. Namely, we express the matrix $Q$ as the product of three matrices: $U$, an orthogonal (by columns) matrix of size $m\times n$, $D$, a diagonal matrix of size $n\times n$, whose entries are greater or equal to zero, and finally, $V$, a transpose matrix of an orthogonal one, of size $n\times n$. That is,
\begin{equation}\label{rc5}
    Q=U\cdot D \cdot V^T.
\end{equation}
Since both $U$ and $V$ have orthogonal columns then $U^TU=V^TV=I$ and $V\cdot V^T=I$, therefore $D$ will be the diagonal matrix given by:
\begin{displaymath}
    D=\left[
    \begin{array}{cccc}
        w_1 & 0 & \ldots & 0 \\
        0 & w_2 & \ldots & 0 \\
        \vdots & \vdots & \ddots & \vdots \\
        0 & 0 & \ldots & w_n
    \end{array}
    \right]\\
\end{displaymath}
where $w_i$ will be the singular values of $Q$ and $w_i\geq 0\ \forall i$.

Having obtained the matrices $U$, $V$ and $D$ we will take $e_1$ and $e_2$ as the first and second column of $V^T$ (which are mutually orthogonal) respectively. The plane formed by these orthogonal vectors is the plane that minimises the sum of the distances of each vector $q_i$ to it.

A basis that generates the plane we need is given by
$$\begin{array}{lll}
    e_1&=&[ -0.2202,-0.4050,-0.1848,0,-0.4050,-0.4050,0,\\
       & & -0.2202,0,-0.3291,-0.4050,-0.2286,-0.2286\ ]
  \end{array}$$
and
$$\begin{array}{lll}
  e_2&=& [-0.5118,0.1032,0.6150,0,0.1032,0.1032,0,\\
 & & -0.5118,0,-0.2273, 0.1032,0.0422,0.0422\ ].
 \end{array}$$

Now we can compute the projection $Pq_i$ of each vector $q_i$  to the plane $\Pi=\Pi<e_1,e_2>$,
\begin{equation}
    Pq_i=(q_i\cdot e_1)e_1+(q_i\cdot e_2)e_2.
\end{equation}
If we take $e_1$ and $e_2$ as the vectors that generate respectively the horizontal and vertical axis of a bidimensional coordinate system, we can compute both coordinates of each vector $q_i$ with respect to $\{e_1,e_2\}$, using the following equation
\begin{equation}
    x_{q_i}=Pq_i\cdot e_1, \ \ \ \ y_{q_i}=Pq_i\cdot e_2
\end{equation}
where $x$ and $y$ denote the horizontal and vertical axis respectively.

\section{Computation of the solution}\label{ap:sol}

The objective is to find the stationary solutions of (\ref{eq:rd1}), that is

\begin{equation}\label{rd2}
\begin{array}{lll}
  d_1\Delta u+f(u,v) & = & 0\\
  d_2\Delta v+g(u,v) & = & 0.
\end{array}
\end{equation}

Given the geometry of the flower, we will consider an anular domain $\Omega$, centered at the origin. That is, a disc of radius $R$, with a concentric circular perforation in the origin of radius $r_\epsilon$. This assumption  simplifies the numerics avoiding a singularity due to the choice of polar coordinates and it is a biologically coherent.
The domain $\Omega$ is thus defined as
\begin{equation}\label{rd2a}
\Omega=\{z:r_\epsilon\leq |z| \leq R\}.
\end{equation}
We will work with polar coordinates $(r,\theta)$, since it is more natural given the geometry of the flower.
Our system (\ref{rd2}), after this coordinate transformation is the following
\begin{equation}\label{rd3}
\begin{array}{l}
  d_1\left(u_{rr}+\tfrac{1}{r}u_r+\tfrac{1}{r^2}u_{\theta\theta}\right)+f(u,v)=0\\
  d_2\left(v_{rr}+\tfrac{1}{r}v_r+\tfrac{1}{r^2}v_{\theta\theta}\right)+g(u,v)=0.
\end{array}\end{equation}
Let $k\in\{S,P,T,C\}$ be such that
$F(u,v)=a_k[(u-u_k)^2+(v-v_k)^2]$, then the gradient of $F$, $(f,g)=-\nabla F$, will be
$$f=-2a_k(u-u_k),\ \ \ \ \ g=-2a_k(v-v_k).$$
Note that $f$ only depends of $u$ and $g$ only on $v$, so that the two equations in (\ref{rd3}) are uncoupled and hence can be solved separately.

Taking into account the geometry of the flower, it is reasonable to restrict the solutions to those that are radially symmetric.
The partial differential equations (\ref{rd3}) are thus reduced to:

\begin{eqnarray}
  d_1 \left[u''+\tfrac{1}{r }u'\right]+f(u)=0 \label{rd5a} \\
  d_2 \left[v''+\tfrac{1}{r} v'\right]+g(v)=0   \label{rd5b}
\end{eqnarray}
where $u', v'$ are the derivatives with respect to $r$ of $u$ and $v$ respectively.
Since both equations are identical (except for the parameter values), it suffices to work with only one of them, we will work with the first one.

\subsubsection{Boundary value problem}

We solve equation (\ref{rd5a}) in the  anular domain $\Omega$ (defined in (\ref{rd2a})), centered at the origin. It is natural to start in a point $(u_0,v_0)$ in the sepal's basin. Also, to be able to work with homogeneous boundary conditions, we do a translation of the domain of $-r_\epsilon$ in the $r$-axis. The new domain $\tilde\Omega$ is defined as:
$$\tilde\Omega=\{z:0 \leq |z-r_\epsilon| \leq R-r_\epsilon \}.$$
We'll require the derivative of the solution in $r=0$ to be zero.

The boundary value problem that we will solve is the following:
\begin{equation}\label{vf1a}
\left\{
   \begin{array}{lll}
   d_1\left[u''+\frac{1}{r}u'\right]+f(u)=0\\
   u(R)=u_0;\ u'(0)=0; \ r\in\tilde\Omega
   \end{array}
   \right.
\end{equation}
where $f=-2a_S(u-u_S)$.
To be able to work with homogeneous boundary values we do the change of variables $\tilde u=u-u_0$. Rewriting equation (\ref{vf1a}) in Sturm-Liouville form, and renaming $\tilde u$ as $u$ (to keep notation simple), we get
\begin{equation}\label{vf3}
 [ru']'+c_1 ru=c_2r
\end{equation}
where $c_1=-2a_S/d_1$ y $c_2=c_1(u_S-u_0)$.
Let $h(r)=c_2r$. We define the differential operator $\LL$ the following way
$$\LL=\frac{d}{dr}\left[r\frac{d}{dr}\right]+c_1r$$
hence (\ref{vf3}) can be rewritten as
\begin{equation}\label{vf4}
    \LL u=h(r),
\end{equation}
where $h(r)\leq c_2r$. We now consider the eigenvalue problem
\begin{equation}\label{vf5}
\LL \phi(r)=-\lambda\phi(r)\sigma(r),
\end{equation}
where $\lambda$ is the eigenvalue corresponding to the eigenfunction $\phi$ and $\sigma$ is a weight function that will be adjusted, subject to the boundary conditions $\phi(R)=0$ and $\phi'(0)=0$.

Suppose that the solution $u(r)$ of (\ref{vf4}), can be rewritten as an eigenfunction expansion of the problem (\ref{vf5}), that is
\begin{equation}\label{vf6}
u(r)=\sum_{n=0}^\I b_n\phi_n(r)
\end{equation}
where
\begin{equation}\label{vf7}
b_m=-\frac{\int_0^{\tilde R} h(r)\phi_m(r) dr}{\lambda_m\int_0^{\tilde R}\phi^2_m(r)\sigma(r)dr}
\end{equation}
(for $m=0,\ldots\I$) are the expansion coefficients that are obtained by substituting (\ref{vf6}) into (\ref{vf4}) and using the orthogonality of the eigenfunctions.

We are now interested in finding explicitly the eigenfunctions and eigenvalues of the problem (\ref{vf5}).

\subsubsection{Sturm-Liouville problem} \label{secSL}

We will search for the solutions of eigenvalue problem (\ref{vf5}). Reorganizing terms, we get
$$r^2\phi''+r\phi'+(c_1r^2+\lambda\sigma r)\phi  =  0.$$
Let $\xi^2=\lambda$ (by S-L theory, we know that all the eigenvalues $\lambda$ are simple and real). To simplify the equation, we let $c_1=1$ which implies that $d_1=-2a_S$, (recall that $d_1$ is a diffusion constant that we want to adjust so that our model works properly) and that $c_2=(u_S-u_0)$. Setting $\sigma=-1/r$, last equation is reduced to
\begin{equation}\label{pb1}
r^2\phi''+r\phi'+(r^2-\xi^2)\phi=0,
\end{equation}
Bessel's equation.

A first solution to Bessel's equation (assuming $\xi>0$) will be
\begin{equation}\label{pb4}
J_\xi(r)=\sum_{n=0}^\I\frac{(-1)^n}{n!\Gamma(\xi+n+1)}\left(\frac{r}{2}\right)^{\xi+2n}
\end{equation}
(Bessel's equation of the first kind, of order $\xi$). Since this equation has no (finite) singular points, except for the origin, the series will converge for all values of $r$ when $\xi\geq 0$.

If $\xi$ is not an integer then the second (linearly independent) solution will be
\begin{equation}\label{pb5}
J_{-\xi}(r)=\sum_{n=0}^\I\frac{(-1)^n}{n!\Gamma(-\xi+n+1)}\left(\frac{r}{2}\right)^{-\xi+2n}
\end{equation}
otherwise it will be
\begin{equation}\label{pb6}
Y_\xi(r)=\frac{J_\xi(r)\cos(\xi\pi)-J_{-\xi(x)}}{\sin(\xi\pi)}.
\end{equation}

The general solution will be
\begin{equation}
 \phi(r)=\left\{
   \begin{array}{lll}
   k_1 J_\xi(r)+k_2J_{-\xi}(r), \mbox{\ \ $\xi$ not integer}\\
   k_1 J_\xi(r)+k_2Y_{\xi}(r),  \mbox{\ \ otherwise}.
   \end{array}
   \right.
\end{equation}
where constants $k_1$ and $k_2$ are to be determined. In both cases, to satisfy boundary conditions we need $k_2=0$, so that the general solution to equation (\ref{pb1}) in any case is
$$\phi(r)=k_1 J_\xi(r).$$
Note that for it to be well defined in $r=0$ we need $\xi>1$.

\subsubsection{Boundary conditions}

We now write $\phi(r)=J_\xi(r)$ (choosing $k_1=1$) in its integral form
$$J_\xi(r)=\frac{1}{\pi}\int_0^\pi cos(r\sin\theta-\xi\theta)d\theta.$$
We want it to satisfy $\phi'(0)=J'_\xi(0)=0$. Differentiating $J_\xi$ with respect to $r$, evaluating at zero and solving for $J_\xi'(0)$ we obtain
$$J_\xi'(0)= \frac{1}{2\pi}\frac{(1+\xi)\sin((1-\xi)\pi)-(1-\xi)\sin((1+\xi)\pi)}{(1-\xi^2)}$$
Now, solving $J_\xi'(0)=0$, we get $0 = 2\sin(\pi \xi) \Leftrightarrow \xi=n \Leftrightarrow \lambda=\xi^2=n^2,\ \ n=2,3,4,\ldots$,
that is, the eigenvalues of problem (\ref{vf5}) are the squares of the natural numbers greater than two, and the corresponding eigenfunctions are
$$\phi_n(r)=J_\xi^{(n)}(r)=\frac{1}{\pi}\int_0^\pi \cos(r\sin\theta-n\theta)d\theta$$
with $n=2,3,\ldots$.

We've solved eigenvalue problem (\ref{vf5}). Solution to problem (\ref{vf1a}) is then
\begin{equation}\label{cf1a}
u(r)=\sum_{n=2}^\I \frac{b_n}{\pi}\int_0^\pi \cos(r\sin\theta-n\theta)d\theta
\end{equation}
with
\begin{equation}\label{cf1b}
 b_n=\frac{(u_S-u_0)\pi}{n^2}\frac{\int_0^{\tilde R} r\left(\int_0^\pi \cos(r\sin\theta-n\theta)d\theta \right) dr}{\int_0^{\tilde R}\frac{1}{r}\left(\int_0^\pi \cos(r\sin\theta-n\theta)d\theta \right)^2dr}.
\end{equation}

The solution to problem with dependent variable is $v$, is obtained analogously. We get
\begin{equation}\label{cf2a}
v(r)=\sum_{n=2}^\I \frac{c_n}{\pi}\int_0^\pi \cos(r\sin\theta-n\theta)d\theta
\end{equation}
with
\begin{equation}\label{cf2b}
c_n=\frac{(v_S-v_0)\pi}{n^2}\frac{\int_0^{\tilde R} r\left(\int_0^\pi \cos(r\sin\theta-n\theta)d\theta \right) dr}{\int_0^{\tilde R}\frac{1}{r}\left(\int_0^\pi \cos(r\sin\theta-n\theta)d\theta \right)^2dr}.
\end{equation}

Now, the only thing left is to compute the value of parameter $R$. For this we use the first boundary condition $(u,v)(\tilde R)=(0,0)$, that is, we look for the value of $\tilde R$ for which the solution (for $u$ and $v$) at that specific value is zero. We solve iteratively the integrals in equations (\ref{cf1a}, \ref{cf1b}) and (\ref{cf2a}, \ref{cf2b}), adjusting the values of $\tilde R$ until we find $\tilde R$ such that $(u,v)(\tilde R)=0$. Sums (\ref{cf1a}) and (\ref{cf2a}) are truncated: we will only use the minimum number of terms necessary of the solution to converge.

Recall that solutions $(u,v)$ where translated by a distance $(u_0,v_0)$ and by a value $-r_\epsilon$ in the $r-$axis. The solution to the original problem(\ref{vf1a})in domain $\Omega$ will be given by
\begin{equation}\label{cf3}
u(r)=u_0+\tilde u(r+r_\epsilon),
\end{equation}
where $\tilde u$ is used for solution (\ref{cf1a}) (to keep $u$ for the solution to the original problem).
Analogously, for the problem whose dependent variable is $v$ we get
\begin{equation}\label{cf4}
v(r)=v_0+\tilde v(r+r_\epsilon).
\end{equation}

\subsubsection{Initial value problem}
\begin{figure}[H]
\centering
  \includegraphics[width=8cm]{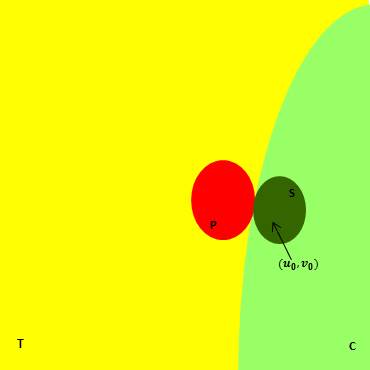}\\
  \caption{Basins of attraction in the epigenetic landscape (potential field $F(u,v)$), each one corresponds to a specific flower organ (S: Sepals, P: Petals, T: Stamens, C: Carpels), $(u_0,v_0)$ is initial condition and it is located in a point inside the basin of sepals.}\label{cuencas}
\end{figure}

The solutions of the boundary value problem obtained in the previous section will be valid only in the interval $I_1=[r_1,R]$, where $r_1\in[r_\epsilon,R)$ is such that $F((u_S,v_S)(r))=P_S$ for all $r\in I_1$ but $F((u_S,v_S)(r_1-\Delta r))\neq P_S$ for a given $\Delta r$. In other words, these solutions will only be valid while we are in the sepal's basin. Once the solution goes out of this basin, parameters of the equations will change, so we will need to compute new set of solutions (see figure \ref{cuencas}). We will call the first couple of solutions (\ref{cf3},\ref{cf4}), $u_S(r)$ and $v_S(r)$.

Suppose that $\tilde r=r_1+\Delta r$ is such that $F((u_S,v_S)(\tilde r))=P_k$, where $k=P,T,C$, and let $\tilde u=u_S(r_1)$, $\tilde v=v_S(r_1)$. We compute
$$\tilde u'=\frac{d}{dr}u_S(r_1)\ \ \ \  \mbox{and}\ \ \ \ \tilde v'=\frac{d}{dr}v_S(r_1).$$

We now have the following initial boundary problems
$$\left\{
   \begin{array}{lll}
   d_1\left[u''+\frac{1}{r}u'\right]+f(u)=0\\
   u(r_1)=\tilde u;\ u'(r_1)=\tilde u'
   \end{array}
   \right.
$$
and
$$\left\{
   \begin{array}{lll}
   d_2\left[v''+\frac{1}{r}v'\right]+g(v)=0\\
   v(r_1)=\tilde v;\ v'(r_1)=\tilde v'
   \end{array}
   \right.
$$
where $f(u)=-2a_k(u-u_k)$, $g(v)=-2a_k(u-u_k)$ (recall that we are in basin $k\in\{P,E,C\}$).

\begin{figure}[H]
\centering
  \includegraphics[width=8cm]{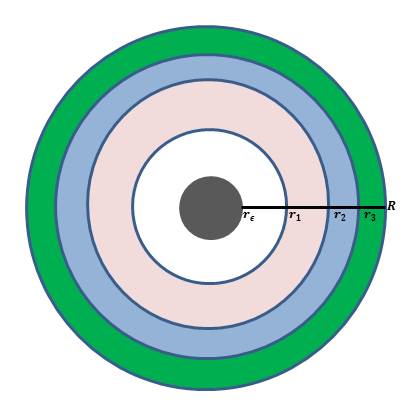}\\
  \caption{Each anular portion of the domain shows where each computed solution is valid: first solution is valid for $r\in[r_\epsilon,r_1)$, second solution is valid for $r\in[r_1,r_2)$, etc.}\label{radios2}
\end{figure}

Again solutions $u=u_k(r)$ and $v=v_k(r)$ found for these initial value problem will only be valid while we are in basin $k$. We now look for the first value $r=r_2$  for which $F((u_k,v_k)(r_2-\Delta r))\neq P_k$. This new solution will be valid for $r\in[r_2,r_1)$, $(r_2<r_1)$.
Plotting $r_2$ into the value of new solutions and derivatives we will get a new set of initial value problems. We continue in this manner until we find a $j-th$ iteration such that $r_h\leq r_\epsilon$, obtaining a set of $j$ solutions each one valid in an anular domain (see fig. \ref{radios2}). \textit{Gluing} the whole set of solutions (smoothly) we will obtain the complete solution to the problem, which by construction, will be continuous and differentiable.

\clearpage
\bibliographystyle{amsplain}
\bibliography{references}

\end{document}